\documentclass[prb,aps,twocolumn,superscriptaddress,floatfix]{revtex4}
\usepackage{graphicx}
\usepackage{amsmath}
\usepackage{amssymb}
\usepackage[linktocpage]{hyperref}

\newcommand{\cgo}{CuGeO$_3$}
\newcommand{\sds}[2]{\vec{S}_{#1}\!\cdot\!\vec{S}_{#2}}

\graphicspath{{.}}

\begin{document}

\title{Phase diagram of the spin-Peierls chain with local coupling}

\author{Alexander Wei{\ss}e}
\affiliation{Institut f{\"u}r Physik, Ernst-Moritz-Arndt-Universit{\"a}t Greifswald, Domstra{\ss}e 10a, 17487 Greifswald, Germany}
\author{Georg Hager}
\affiliation{Regionales Rechenzentrum Erlangen, HPC Services, Martensstra{\ss}e 1, 91058 Erlangen, Germany}
\author{Alan R. Bishop}
\affiliation{Los Alamos National Laboratory, Los Alamos, New Mexico 87545, USA}
\author{Holger Fehske}
\affiliation{Institut f{\"u}r Physik, Ernst-Moritz-Arndt-Universit{\"a}t Greifswald, Domstra{\ss}e 10a, 17487 Greifswald, Germany}

\date{\today}

\begin{abstract}
  We explore the ground-state phase diagram of a Heisenberg spin chain
  coupled locally to optical phonons (bond coupling), using
  large-scale density matrix renormalization group calculations and an
  extended perturbative analysis. For the quantum phase transition
  from the spin liquid to the dimerized phase, we find deviations from
  previous quantum Monte Carlo and flow equation results.
\end{abstract}


\maketitle


The interaction of electronic and lattice degrees of freedom in
combination with reduced dimensionality can lead to a variety of
interesting effects, one of which is the instability of a
one-dimensional metal towards lattice distortion and the opening of a
gap at the Fermi surface that was first described by
Peierls~\cite{Pe55}. A similar effect is observed in quantum spin
chains, where the coupling to the lattice can cause a transition from
a spin liquid with gap-less excitations to a dimerized phase with an
excitation gap.  Experimentally such behavior was first observed in
the 1970s for organic compounds of the TTF and TCNQ
family~\cite{Brea75}, but the topic regained attention after the
discovery of the first inorganic spin-Peierls compound \cgo{} in 1993
by~\textcite{HTU93}. In this material Cu$^{2+}$ ions form well
separated spin-$1/2$ chains with an exchange interaction that couples
to high-frequency \emph{optical} phonons ($\omega\approx J$), and the
phonon dynamics at the phase transition is governed by a central peak
rather than a soft-mode behavior.~\cite{BHRDR98,FHW00} These features
distinguish \cgo{} from other spin-Peierls systems and sparked the
interest in a non-adiabatic modelling.

A good starting point is the study of simplified microscopic models,
which can be build from three ingredients,
\begin{equation}
  H = H_s + H_p + H_{sp}\,.
\end{equation}
Here $H_s = \sum_i \sds{i}{i+1}$ and $H_p = \omega \sum_i
b_i^{\dagger} b_i^{}$ describe a Heisenberg spin-$1/2$ chain and a set
of harmonic (Einstein) oscillators which are coupled by an interaction
term $H_{sp}$. For this interaction we can consider two simple forms,
\begin{align}
  H_{sp}^{\text{diff}} & = \smash[b]{g\omega \sum_i (b_i^{\dagger} + b_i^{}) 
  (\sds{i}{i+1}-\sds{i-1}{i})\,,}\\
  \intertext{and}
  H_{sp}^{\text{loc}} & = \smash[t]{g\omega \sum_i (b_i^{\dagger} + b_i^{}) \sds{i}{i+1}\,.}
\end{align}

The first type of spin-phonon interaction, $H_{sp}^{\text{diff}}$, has
been studied with a number of methods, including perturbation
theory~\cite{KF87,WWF99}, flow equations~\cite{Uh98,RBU01}, exact
diagonalization~\cite{WWF99} and DMRG~\cite{BMH99}. The latter approach
identified the ground-state phase diagram, but also analytically the
quantum phase transition from the gap-less to the dimerized phase is
rather well understood: For finite phonon frequency $\omega$ the
spin-phonon coupling $g$ leads to effective spin interactions beyond
nearest-neighbor exchange, i.e., the low energy physics is governed by
a frustrated Heisenberg model.  As we know from the spin model
\begin{equation}\label{frust}
  H = \sum_i (\sds{i}{i+1} + \alpha\sds{i}{i+2})\,,
\end{equation}
frustration can lead to dimerization if the parameter $\alpha$ exceeds
a certain critical value ($\alpha_c = 0.241167$ in this
case~\cite{ON92,CCE95,Eg96}), and similarly we obtain a finite
$g_c(\omega)$.\cite{Uh98,BMH99,WWF99,HWWJF06p}

\begin{figure}
  \includegraphics[width=\linewidth]{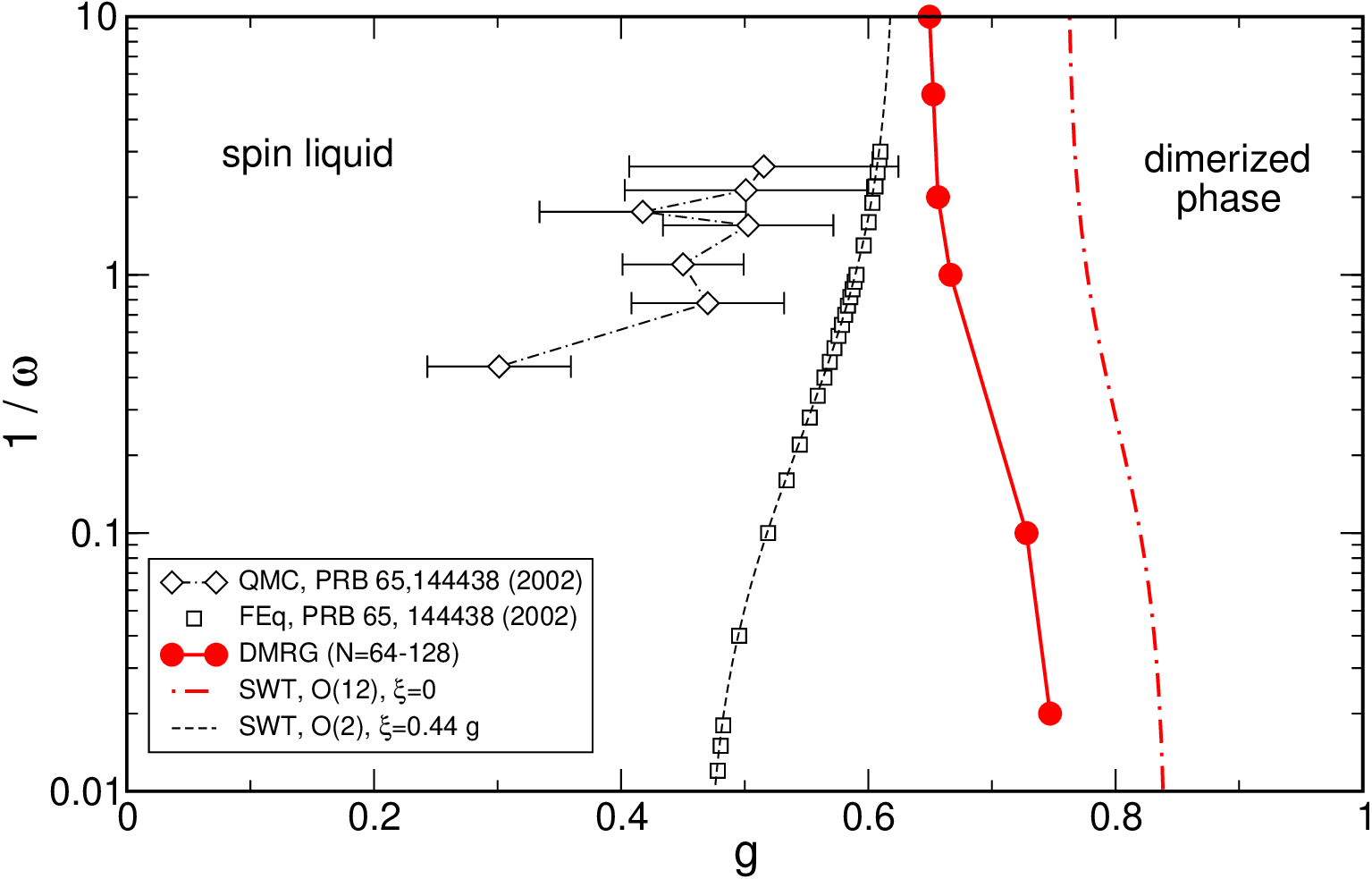}
  \caption{(Color online) Ground-state phase diagram of the
    spin-Peierls chain with local coupling: QMC and flow equation
    (FEq) results of \textcite{RLUK02} compared to DMRG data and
    results of 12th order Schrieffer-Wolff transformation
    (SWT).}\label{fig:pd}
\end{figure}
For the second type of spin-phonon coupling, $H_{sp}^{\text{loc}}$,
which applies to \cgo{}, to date the precise location of the phase
boundary was arguable. In previous studies~\cite{WWF99,WFK98} we
calculated $g_c$ using perturbation theory, a variational ansatz, as
well as exact diagonalization of small systems. This was challenged by
results of the flow equation method and, in a limited parameter range,
by quantum Monte Carlo.\cite{RLUK02} In this article we present
unbiased results of large-scale density matrix renormalization group
(DMRG) calculations, and extend our perturbation theory by several
orders in $g$, high enough to ensure convergence. The new results are
summarized in Figure~\ref{fig:pd} and compared to other approaches.
More details now follow.

\paragraph*{Numerical results:} DMRG calculations are certainly the most
precise numerical tool for studying the low energy properties of
one-dimensional models, but as yet have only been performed for the
spin-Peierls model with difference coupling~\cite{BMH99}
$H_{sp}^{\text{diff}}$, models with acoustic phonons~\cite{BB05}, or
with $XY$-type spin interaction~\cite{CM96}. We therefore implemented
the local spin-phonon interaction~$H_{sp}^{\text{loc}}$, using a
high-performance, parallel version of the usual two-block
finite-lattice algorithm with up to 1000 states per block. To
detect the quantum phase transition from the gap-less to the dimerized
phase we use the established criterion of the level-crossing between
the first singlet and the first triplet excitation, which was derived
for the frustrated spin chain~\cite{AGSZ89}, Eq.~\eqref{frust}, and
has successfully been applied~\cite{BMH99,HWWJF06p} in the case of
$H_{sp}^{\text{diff}}$. For finite systems in the gap-less phase the
lowest singlet excitation is above the lowest triplet, both becoming
degenerate with the singlet ground state for system size
$N\to\infty$. In the gapped phase, for $N\to\infty$ the lowest singlet
becomes degenerate with the ground state to form the symmetry-broken
dimerized state, whereas the lowest triplet maintains a finite gap.
Consequently, the two excitations will cross at the critical point.
Note however, that for small phonon frequency $\omega$ the relevant
singlet excitation can be confused with a copy of the ground state
plus an excited phonon. This frequency range therefore requires rather
large $N$ for the correct crossing to be detectable in the low energy
spectrum of the spin-Peierls models.

\begin{figure}
  \includegraphics[width=\linewidth]{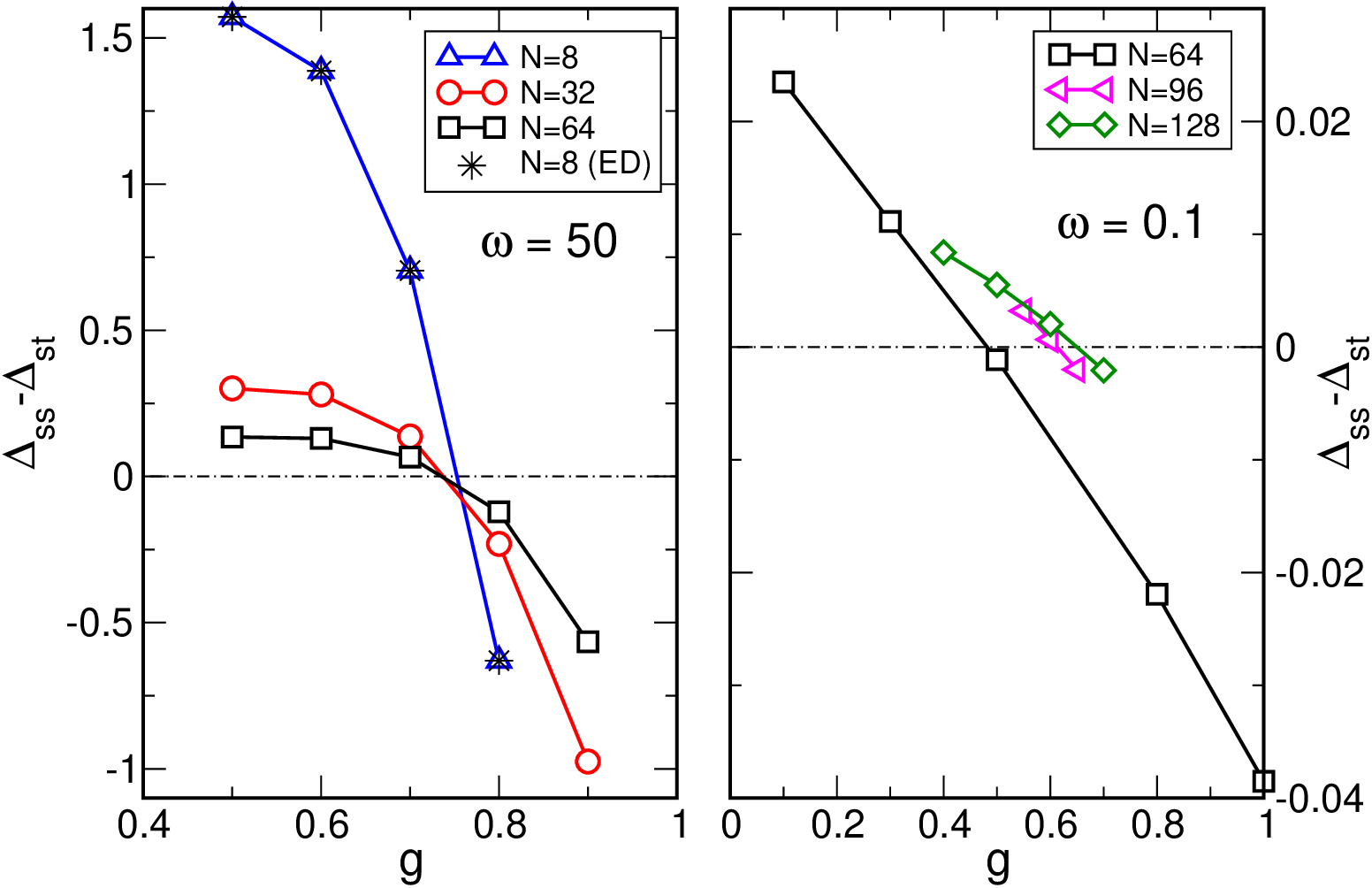}
  \caption{(Color online) Crossing of the singlet and triplet gaps for
    large (right) and small (left) phonon frequency. Note the
    different finite-size scalings.}\label{fig:cr}
\end{figure}

In Figure~\ref{fig:cr} we show the level difference as a function of
the spin-phonon coupling for various system sizes and two typical
phonon frequencies. For $\omega \gtrsim 0.5$ there is almost no
finite-size dependence of the critical coupling $g_c$, whereas for
smaller frequencies the data scales noticeably. It is tempting to
attribute this different behavior to the cross-over from the
anti-adiabatic to the adiabatic regime, na\"ively expected for
$\omega\sim J \equiv 1$. However, as was pointed out by
\textcite{COG05} using bosonization techniques, the relevant scale for
the adiabatic to anti-adiabatic cross-over is given by the excitation
gap, $\omega\sim\Delta$.  The larger $N$-dependence of $g_c$ observed
in Figure~\ref{fig:cr} for $\omega=0.1$ is therefore related to the
finite-size gap still being close $\omega$, and will disappear for
$N\to\infty$. Note also, that the bosonization results support the
analytical approach we present in the following, which is based on the
assumption of anti-adiabaticity.


\begin{figure}
  \includegraphics[width=\linewidth]{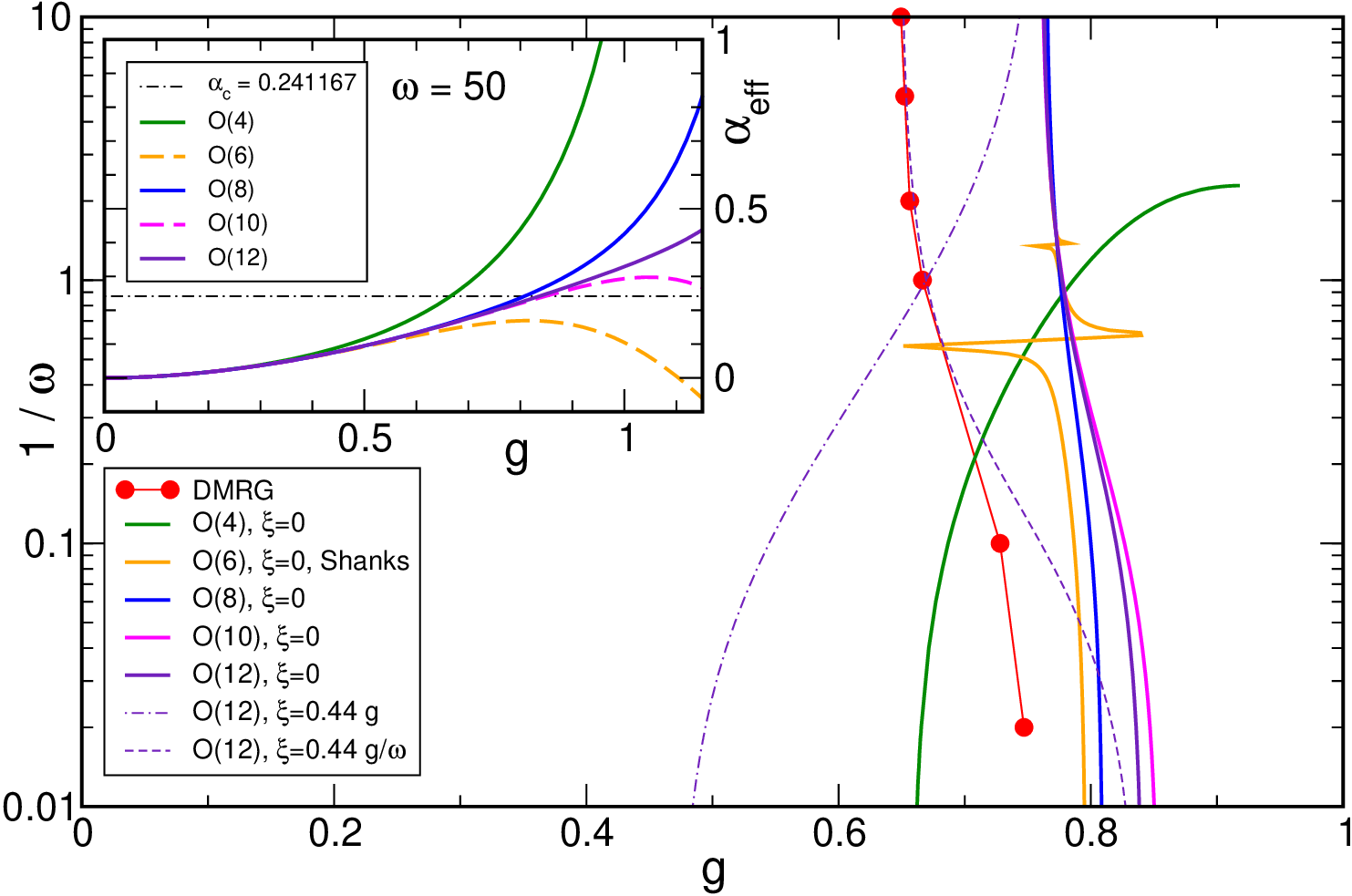}
  \caption{(Color online) Convergence of the Schrieffer-Wolff approach
    with increasing expansion order $O(k)$. Main panel: phase diagram showing
    DMRG and expansion data for order $4$--$12$ without and with
    oscillator shifts. Inset: $\alpha_{\text{eff}}$ as a function of
    $g$ for $\omega=50$ and orders $4$--$12$.}\label{fig:cv}
\end{figure}

\begin{table*}
  \caption{Expansion of the effective long-ranged exchange 
    $J_n \sds{i}{i+n}$ in powers of the spin-phonon 
    coupling $g$.}\label{tab:jn}
  \renewcommand{\arraystretch}{1.2}
  \begin{tabular}{l|cccccccc}
    & $ J_0 $ & $ J_1 $ & $ J_2 $ & $ J_3 $ & $ J_4 $ & $ J_5 $ & $ J_6 $ & $ J_7 $ \\
    \hline
    $    g^0 $ & $ \xi^2 \omega $ & $ 1 $ & $ \cdot $ & $ \cdot $ & $ \cdot $ & $ \cdot $ & $ \cdot $ & $ \cdot $ \\
    $    g^2 $ & $ \cdot $ & $ -\frac{1}{2} $ & $ \frac{1}{2} $ & $ \cdot $ & $ \cdot $ & $ \cdot $ & $ \cdot $ & $ \cdot $ \\
    $    g^2 \omega $ & $ -\frac{3}{16} $ & $ \frac{1}{2} $ & $ \cdot $ & $ \cdot $ & $ \cdot $ & $ \cdot $ & $ \cdot $ & $ \cdot $ \\
    $    g^3 \xi \omega $ & $ \cdot $ & $ -\frac{2}{3} $ & $ \frac{2}{3} $ & $ \cdot $ & $ \cdot $ & $ \cdot $ & $ \cdot $ & $ \cdot $ \\
    $    g^4 $ & $ \cdot $ & $ \frac{7}{24} $ & $ -\frac{37}{96} $ & $ \frac{3}{32} $ & $ \cdot $ & $ \cdot $ & $ \cdot $ & $ \cdot $ \\
    $    g^4 \omega $ & $ \frac{3}{64} $ & $ -\frac{3}{8} $ & $ \frac{3}{16} $ & $ \cdot $ & $ \cdot $ & $ \cdot $ & $ \cdot $ & $ \cdot $ \\
    $    g^5 \xi \omega $ & $ \cdot $ & $ \frac{7}{15} $ & $ -\frac{37}{60} $ & $ \frac{3}{20} $ & $ \cdot $ & $ \cdot $ & $ \cdot $ & $ \cdot $ \\
    $    g^6 $ & $ \cdot $ & $ -\frac{697}{5760} $ & $ \frac{541}{2880} $ & $ -\frac{29}{384} $ & $ \frac{5}{576} $ & $ \cdot $ & $ \cdot $ & $ \cdot $ \\
    $    g^6 \omega $ & $ -\frac{1}{64} $ & $ \frac{29}{144} $ & $ -\frac{91}{576} $ & $ \frac{5}{192} $ & $ \cdot $ & $ \cdot $ & $ \cdot $ & $ \cdot $ \\
    $    g^7 \xi \omega $ & $ \cdot $ & $ -\frac{697}{3360} $ & $ \frac{541}{1680} $ & $ -\frac{29}{224} $ & $ \frac{5}{336} $ & $ \cdot $ & $ \cdot $ & $ \cdot $ \\
    $    g^8 $ & $ \cdot $ & $ \frac{5119}{129024} $ & $ -\frac{88339}{1290240} $ & $ \frac{45749}{1290240} $ & $ -\frac{3685}{516096} $ & $ \frac{35}{73728} $ & $ \cdot $ & $ \cdot $ \\
    $    g^8 \omega $ & $ \frac{107}{24576} $ & $ -\frac{3611}{46080} $ & $ \frac{3569}{46080} $ & $ -\frac{683}{30720} $ & $ \frac{35}{18432} $ & $ \cdot $ & $ \cdot $ & $ \cdot $ \\
    $    g^9 \xi \omega $ & $ \cdot $ & $ \frac{5119}{72576} $ & $ -\frac{88339}{725760} $ & $ \frac{45749}{725760} $ & $ -\frac{3685}{290304} $ & $ \frac{35}{41472} $ & $ \cdot $ & $ \cdot $ \\
    $    g^{10} $ & $ \cdot $ & $ -\frac{2516207}{232243200} $ & $ \frac{9363217}{464486400} $ & $ -\frac{105569}{8601600} $ & $ \frac{171601}{51609600} $ & $ -\frac{7601}{19353600} $ & $ \frac{7}{409600} $ & $ \cdot $ \\
    $    g^{10} \omega $ & $ -\frac{18101}{17203200} $ & $ \frac{35017}{1433600} $ & $ -\frac{72439}{2580480} $ & $ \frac{1243}{115200} $ & $ -\frac{42433}{25804800} $ & $ \frac{7}{81920} $ & $ \cdot $ & $ \cdot $ \\
    $    g^{11} \xi \omega $ & $ \cdot $ & $ -\frac{2516207}{127733760} $ & $ \frac{9363217}{255467520} $ & $ -\frac{105569}{4730880} $ & $ \frac{171601}{28385280} $ & $ -\frac{691}{967680} $ & $ \frac{7}{225280} $ & $ \cdot $ \\
    $    g^{12} $ & $ \cdot $ & $ \frac{624432139}{245248819200} $ & $ -\frac{820409053}{163499212800} $ & $ \frac{62408713}{18166579200} $ & $ -\frac{138813341}{122624409600} $ & $ \frac{12753401}{70071091200} $ & $
    -\frac{6906257}{490497638400} $ & $ \frac{77}{176947200} $ \\
    $    g^{12} \omega $ & $ \frac{69371}{309657600} $ & $ -\frac{441857}{68812800} $ & $ \frac{1701589}{206438400} $ & $ -\frac{7128059}{1857945600} $ & $ \frac{280187}{348364800} $ & $ -\frac{4939}{66355200} $ & $
    \frac{77}{29491200} $ & $ \cdot $ \\
    $    g^{13} \xi \omega $ & $ \cdot $ & $ \frac{624432139}{122624409600} $ & $ -\frac{820409053}{81749606400} $ & $ \frac{62408713}{9083289600} $ & $ -\frac{138813341}{61312204800} $ & $ \frac{12753401}{35035545600} $ & $
    -\frac{6906257}{245248819200} $ & $ \frac{77}{88473600} $
  \end{tabular}
\end{table*}

\paragraph*{Analytical results:} Already in our earlier work we suggested
construction of an effective spin model for the low energy physics of the
spin-Peierls problem by integrating out the phonons with a
Schrieffer-Wolff transformation~\cite{SW66} (which removes direct
spin-phonon interactions) followed by an average over the phonon
vacuum. In more detail, we apply the unitary transformation $\tilde H
= \exp(S) H \exp(-S)$, where
\begin{equation}
  S = g \sum_i (b_i^\dagger - b_i)\sds{i}{i+1}\,.
\end{equation}
Unfortunately, this transformation cannot be evaluated exactly, but
needs to be approximated by an expansion in iterated commutators:
\begin{equation}
  \tilde H = \sum_k [S,H]_k / k!,
\end{equation}
where $[S,H]_{k+1} = [S,[S,H]_k]$ and $[S,H]_0 = H$. For increasing
expansion order these commutators quickly become very complicated and
easily involve millions of terms. Using FORM~\cite{Ver00,Ver06}, an
algebra tool popular in high energy physics, we are now able to push
the limit of the expansion to order $k=12$, a tremendous advantage
over our previous results with $k=4$. For the last step, we decided to be
more general by averaging the transformed Hamiltonian over coherent
states with $b_i|\xi\rangle = \xi|\xi\rangle$ instead of just the
phonon vacuum:
\begin{equation}\label{heff}
  \begin{aligned}
    H_{\text{eff}} & = \langle \xi|\tilde H|\xi\rangle \\
    & = J_0 N + \sum_i\sum_{n=1}^{7} J_n \sds{i}{i+n} + \text{multi-spin terms}
  \end{aligned}
\end{equation}
This allows for a direct comparison with the flow equation result of
\textcite{RLUK02}, which is equivalent to a 2nd order expansion
($k=2$) with phonon shift $\xi = - \langle \sds{i}{i+1} \rangle g
\approx (1/4 - \ln 2) g \approx 0.44 g$. Here the spin correlator is
approximated by its value in the isotropic Heisenberg chain. For a
comparison see the small squares and the thin dashed line in
Figure~\ref{fig:pd}.

In Table~\ref{tab:jn} we list the expansion coefficients of the
resulting long-ranged exchange interactions $J_n$ that contribute to
our effective Hamiltonian Eq.~\eqref{heff}. We neglect interactions
that involve more than two spin operators. The phase transition line
in Figure~\ref{fig:pd} is obtained by equating the effective
frustration $\alpha_{\text{eff}} := J_2/J_1$ with the critical value
$\alpha_c$ of the next-nearest-neighbor spin chain (see
Eq.~\eqref{frust}), where for the phonon shift we use the ``neutral''
value $\xi=0$, i.e., the vacuum. Except for a constant offset, the
analytical result matches the DMRG data quite well. In particular, the
slope of the critical line is captured correctly.

The convergence of our analytical approach is illustrated in
Figure~\ref{fig:cv}, depicting the phase transition lines obtained for
increasing expansion order. We observe an oscillatory behavior, such
that reliable data results only beyond order $k\ge 8$. For lower
orders, $\alpha_{\text{eff}}$ may not even reach the critical value,
as is evident from the order $6$ data shown in the inset.
Nevertheless, for order $6$ the phase transition can be estimated
using a Shanks transformation~\cite{BO99}. 

We also checked, if the phonon shift $\xi\approx 0.44 g$ proposed by
\textcite{RLUK02} leads to an improved effective description, but
clearly the dot-dashed line in Figure~\ref{fig:cv} deviates
qualitatively from the DMRG data.  Within the flow equation approach
this shift was motivated by the technical requirement for normal
ordering such that the phonons couple to $(\sds{i}{i+1} -
\langle\sds{i}{i+1}\rangle)$, but not by, e.g., some variational
principle. Hence, the discrepancy is not too surprising.
Interestingly, for $\omega\lesssim 3$ we obtain an almost perfect
description of the phase transition, if instead we assume a phonon
shift of $\xi\approx 0.44 g/\omega$. As yet we did not find a
reasonable physical motivation for this $\omega$-dependence, and the
good agreement might be accidental.


To summarize, using DMRG we obtained the, to date, most precise
numerical result for the location of the quantum phase transition
from the spin liquid to the dimerized phase in the one-dimensional
Heisenberg model with local coupling to optical phonons. In addition,
we proved the convergence of the unitary transformation approach that
maps the full spin-phonon model to an effective frustrated spin model
and allows an analytical calculation of the phase boundary in good
agreement with the numerical data. 

We thank E. Jeckelmann and G. Wellein for many helpful comments. In
addition, we acknowledge the generous grant of resources by HLRN
and NERSC, and financial support by DFG through SPP 1073. Work at
Los Alamos was performed under the auspices of the US DOE.


\begin{thebibliography}{24}
\expandafter\ifx\csname natexlab\endcsname\relax\def\natexlab#1{#1}\fi
\expandafter\ifx\csname bibnamefont\endcsname\relax
  \def\bibnamefont#1{#1}\fi
\expandafter\ifx\csname bibfnamefont\endcsname\relax
  \def\bibfnamefont#1{#1}\fi
\expandafter\ifx\csname citenamefont\endcsname\relax
  \def\citenamefont#1{#1}\fi
\expandafter\ifx\csname url\endcsname\relax
  \def\url#1{\texttt{#1}}\fi
\expandafter\ifx\csname urlprefix\endcsname\relax\def\urlprefix{URL }\fi
\providecommand{\bibinfo}[2]{#2}
\providecommand{\eprint}[2][]{\url{#2}}

\bibitem[{\citenamefont{Peierls}(1955)}]{Pe55}
\bibinfo{author}{\bibfnamefont{R.}~\bibnamefont{Peierls}},
  \emph{\bibinfo{title}{Quantum theory of solids}} (\bibinfo{publisher}{Oxford
  University Press}, \bibinfo{address}{Oxford}, \bibinfo{year}{1955}).

\bibitem[{\citenamefont{Bray et~al.}(1975)\citenamefont{Bray, Hart, Interrante,
  Jacobs, Kasper, Watkins, Wee, and Bonner}}]{Brea75}
\bibinfo{author}{\bibfnamefont{J.~W.} \bibnamefont{Bray}},
  \bibinfo{author}{\bibfnamefont{H.~R.} \bibnamefont{Hart},
  \bibfnamefont{Jr.}}, \bibinfo{author}{\bibfnamefont{L.~V.}
  \bibnamefont{Interrante}}, \bibinfo{author}{\bibfnamefont{I.~S.}
  \bibnamefont{Jacobs}}, \bibinfo{author}{\bibfnamefont{J.~S.}
  \bibnamefont{Kasper}}, \bibinfo{author}{\bibfnamefont{G.~D.}
  \bibnamefont{Watkins}}, \bibinfo{author}{\bibfnamefont{S.~H.}
  \bibnamefont{Wee}}, \bibnamefont{and} \bibinfo{author}{\bibfnamefont{J.~C.}
  \bibnamefont{Bonner}}, \bibinfo{journal}{Phys. Rev. Lett.}
  \textbf{\bibinfo{volume}{35}}, \bibinfo{pages}{744} (\bibinfo{year}{1975}).

\bibitem[{\citenamefont{Hase et~al.}(1993)\citenamefont{Hase, Terasaki, and
  Uchinokura}}]{HTU93}
\bibinfo{author}{\bibfnamefont{M.}~\bibnamefont{Hase}},
  \bibinfo{author}{\bibfnamefont{I.}~\bibnamefont{Terasaki}}, \bibnamefont{and}
  \bibinfo{author}{\bibfnamefont{K.}~\bibnamefont{Uchinokura}},
  \bibinfo{journal}{Phys. Rev. Lett.} \textbf{\bibinfo{volume}{70}},
  \bibinfo{pages}{3651} (\bibinfo{year}{1993}).

\bibitem[{\citenamefont{Braden et~al.}(1998)\citenamefont{Braden, Hennion,
  Reichardt, Dhalenne, and Revcolevschi}}]{BHRDR98}
\bibinfo{author}{\bibfnamefont{M.}~\bibnamefont{Braden}},
  \bibinfo{author}{\bibfnamefont{B.}~\bibnamefont{Hennion}},
  \bibinfo{author}{\bibfnamefont{W.}~\bibnamefont{Reichardt}},
  \bibinfo{author}{\bibfnamefont{G.}~\bibnamefont{Dhalenne}}, \bibnamefont{and}
  \bibinfo{author}{\bibfnamefont{A.}~\bibnamefont{Revcolevschi}},
  \bibinfo{journal}{Phys. Rev. Lett.} \textbf{\bibinfo{volume}{80}},
  \bibinfo{pages}{3634} (\bibinfo{year}{1998}).

\bibitem[{\citenamefont{Fehske et~al.}(2000)\citenamefont{Fehske, Holicki, and
  Wei{\ss}e}}]{FHW00}
\bibinfo{author}{\bibfnamefont{H.}~\bibnamefont{Fehske}},
  \bibinfo{author}{\bibfnamefont{M.}~\bibnamefont{Holicki}}, \bibnamefont{and}
  \bibinfo{author}{\bibfnamefont{A.}~\bibnamefont{Wei{\ss}e}}, in
  \emph{\bibinfo{booktitle}{Advances in Solid State Physics 40}}, edited by
  \bibinfo{editor}{\bibfnamefont{B.}~\bibnamefont{Kramer}}
  (\bibinfo{publisher}{Vieweg}, \bibinfo{address}{Wiesbaden},
  \bibinfo{year}{2000}), pp. \bibinfo{pages}{235--249}.

\bibitem[{\citenamefont{Kuboki and Fukuyama}(1987)}]{KF87}
\bibinfo{author}{\bibfnamefont{K.}~\bibnamefont{Kuboki}} \bibnamefont{and}
  \bibinfo{author}{\bibfnamefont{H.}~\bibnamefont{Fukuyama}},
  \bibinfo{journal}{J. Phys. Soc. Jpn.} \textbf{\bibinfo{volume}{56}},
  \bibinfo{pages}{3126} (\bibinfo{year}{1987}).

\bibitem[{\citenamefont{Wei{\ss}e et~al.}(1999)\citenamefont{Wei{\ss}e,
  Wellein, and Fehske}}]{WWF99}
\bibinfo{author}{\bibfnamefont{A.}~\bibnamefont{Wei{\ss}e}},
  \bibinfo{author}{\bibfnamefont{G.}~\bibnamefont{Wellein}}, \bibnamefont{and}
  \bibinfo{author}{\bibfnamefont{H.}~\bibnamefont{Fehske}},
  \bibinfo{journal}{Phys. Rev. B} \textbf{\bibinfo{volume}{60}},
  \bibinfo{pages}{6566} (\bibinfo{year}{1999}).

\bibitem[{\citenamefont{Uhrig}(1998)}]{Uh98}
\bibinfo{author}{\bibfnamefont{G.~S.} \bibnamefont{Uhrig}},
  \bibinfo{journal}{Phys. Rev. B} \textbf{\bibinfo{volume}{57}},
  \bibinfo{pages}{R14004} (\bibinfo{year}{1998}).

\bibitem[{\citenamefont{Raas et~al.}(2001)\citenamefont{Raas, B{\"u}hler, and
  Uhrig}}]{RBU01}
\bibinfo{author}{\bibfnamefont{C.}~\bibnamefont{Raas}},
  \bibinfo{author}{\bibfnamefont{A.}~\bibnamefont{B{\"u}hler}},
  \bibnamefont{and} \bibinfo{author}{\bibfnamefont{G.~S.} \bibnamefont{Uhrig}},
  \bibinfo{journal}{Eur. Phys. J. B} \textbf{\bibinfo{volume}{21}},
  \bibinfo{pages}{369­} (\bibinfo{year}{2001}).

\bibitem[{\citenamefont{Bursill et~al.}(1999)\citenamefont{Bursill, McKenzie,
  and Hamer}}]{BMH99}
\bibinfo{author}{\bibfnamefont{R.~J.} \bibnamefont{Bursill}},
  \bibinfo{author}{\bibfnamefont{R.~H.} \bibnamefont{McKenzie}},
  \bibnamefont{and} \bibinfo{author}{\bibfnamefont{C.~J.} \bibnamefont{Hamer}},
  \bibinfo{journal}{Phys. Rev. Lett.} \textbf{\bibinfo{volume}{83}},
  \bibinfo{pages}{408} (\bibinfo{year}{1999}).

\bibitem[{\citenamefont{Okamoto and Nomura}(1992)}]{ON92}
\bibinfo{author}{\bibfnamefont{K.}~\bibnamefont{Okamoto}} \bibnamefont{and}
  \bibinfo{author}{\bibfnamefont{K.}~\bibnamefont{Nomura}},
  \bibinfo{journal}{Phys. Lett. A} \textbf{\bibinfo{volume}{169}},
  \bibinfo{pages}{433} (\bibinfo{year}{1992}).

\bibitem[{\citenamefont{Castilla et~al.}(1995)\citenamefont{Castilla,
  Chakravarty, and Emery}}]{CCE95}
\bibinfo{author}{\bibfnamefont{G.}~\bibnamefont{Castilla}},
  \bibinfo{author}{\bibfnamefont{S.}~\bibnamefont{Chakravarty}},
  \bibnamefont{and} \bibinfo{author}{\bibfnamefont{V.~J.} \bibnamefont{Emery}},
  \bibinfo{journal}{Phys. Rev. Lett.} \textbf{\bibinfo{volume}{75}},
  \bibinfo{pages}{1823} (\bibinfo{year}{1995}).

\bibitem[{\citenamefont{Eggert}(1996)}]{Eg96}
\bibinfo{author}{\bibfnamefont{S.}~\bibnamefont{Eggert}},
  \bibinfo{journal}{Phys. Rev. B} \textbf{\bibinfo{volume}{54}},
  \bibinfo{pages}{R9612} (\bibinfo{year}{1996}).

\bibitem[{\citenamefont{Hager et~al.}(2006)\citenamefont{Hager, Wei{\ss}e,
  Wellein, Jeckelmann, and Fehske}}]{HWWJF06p}
\bibinfo{author}{\bibfnamefont{G.}~\bibnamefont{Hager}},
  \bibinfo{author}{\bibfnamefont{A.}~\bibnamefont{Wei{\ss}e}},
  \bibinfo{author}{\bibfnamefont{G.}~\bibnamefont{Wellein}},
  \bibinfo{author}{\bibfnamefont{E.}~\bibnamefont{Jeckelmann}},
  \bibnamefont{and} \bibinfo{author}{\bibfnamefont{H.}~\bibnamefont{Fehske}}
  (\bibinfo{year}{2006}), \bibinfo{note}{preprint},
  \urlprefix\url{http://arXiv.org/abs/cond-mat/0606360}.

\bibitem[{\citenamefont{Raas et~al.}(2002)\citenamefont{Raas, L{\"o}w, Uhrig,
  and K{\"u}hne}}]{RLUK02}
\bibinfo{author}{\bibfnamefont{C.}~\bibnamefont{Raas}},
  \bibinfo{author}{\bibfnamefont{U.}~\bibnamefont{L{\"o}w}},
  \bibinfo{author}{\bibfnamefont{G.~S.} \bibnamefont{Uhrig}}, \bibnamefont{and}
  \bibinfo{author}{\bibfnamefont{R.~W.} \bibnamefont{K{\"u}hne}},
  \bibinfo{journal}{Phys. Rev. B} \textbf{\bibinfo{volume}{65}},
  \bibinfo{pages}{144438} (\bibinfo{year}{2002}).

\bibitem[{\citenamefont{Wellein et~al.}(1998)\citenamefont{Wellein, Fehske, and
  Kampf}}]{WFK98}
\bibinfo{author}{\bibfnamefont{G.}~\bibnamefont{Wellein}},
  \bibinfo{author}{\bibfnamefont{H.}~\bibnamefont{Fehske}}, \bibnamefont{and}
  \bibinfo{author}{\bibfnamefont{A.~P.} \bibnamefont{Kampf}},
  \bibinfo{journal}{Phys. Rev. Lett.} \textbf{\bibinfo{volume}{81}},
  \bibinfo{pages}{3956} (\bibinfo{year}{1998}).

\bibitem[{\citenamefont{Barford and Bursill}(2005)}]{BB05}
\bibinfo{author}{\bibfnamefont{W.}~\bibnamefont{Barford}} \bibnamefont{and}
  \bibinfo{author}{\bibfnamefont{R.~J.} \bibnamefont{Bursill}},
  \bibinfo{journal}{Phys. Rev. Lett.} \textbf{\bibinfo{volume}{95}},
  \bibinfo{pages}{137207} (\bibinfo{year}{2005}).

\bibitem[{\citenamefont{Caron and Moukouri}(1996)}]{CM96}
\bibinfo{author}{\bibfnamefont{L.~G.} \bibnamefont{Caron}} \bibnamefont{and}
  \bibinfo{author}{\bibfnamefont{S.}~\bibnamefont{Moukouri}},
  \bibinfo{journal}{Phys. Rev. Lett.} \textbf{\bibinfo{volume}{76}},
  \bibinfo{pages}{4050} (\bibinfo{year}{1996}).

\bibitem[{\citenamefont{Affleck et~al.}(1989)\citenamefont{Affleck, Gepner,
  Schulz, and Ziman}}]{AGSZ89}
\bibinfo{author}{\bibfnamefont{I.}~\bibnamefont{Affleck}},
  \bibinfo{author}{\bibfnamefont{D.}~\bibnamefont{Gepner}},
  \bibinfo{author}{\bibfnamefont{H.~J.} \bibnamefont{Schulz}},
  \bibnamefont{and} \bibinfo{author}{\bibfnamefont{T.}~\bibnamefont{Ziman}},
  \bibinfo{journal}{J. Phys. A} \textbf{\bibinfo{volume}{22}},
  \bibinfo{pages}{511} (\bibinfo{year}{1989}).

\bibitem[{\citenamefont{Citro et~al.}(2005)\citenamefont{Citro, Orignac, and
  Giamarchi}}]{COG05}
\bibinfo{author}{\bibfnamefont{R.}~\bibnamefont{Citro}},
  \bibinfo{author}{\bibfnamefont{E.}~\bibnamefont{Orignac}}, \bibnamefont{and}
  \bibinfo{author}{\bibfnamefont{T.}~\bibnamefont{Giamarchi}},
  \bibinfo{journal}{Phys. Rev. B} \textbf{\bibinfo{volume}{72}},
  \bibinfo{pages}{024434} (\bibinfo{year}{2005}).

\bibitem[{\citenamefont{Schrieffer and Wolff}(1966)}]{SW66}
\bibinfo{author}{\bibfnamefont{J.~R.} \bibnamefont{Schrieffer}}
  \bibnamefont{and} \bibinfo{author}{\bibfnamefont{P.~A.} \bibnamefont{Wolff}},
  \bibinfo{journal}{Phys. Rev.} \textbf{\bibinfo{volume}{149}},
  \bibinfo{pages}{491} (\bibinfo{year}{1966}).

\bibitem[{\citenamefont{Vermaseren}(2000)}]{Ver00}
\bibinfo{author}{\bibfnamefont{J.~A.~M.} \bibnamefont{Vermaseren}}
  (\bibinfo{year}{2000}), \bibinfo{note}{preprint},
  \urlprefix\url{http://arXiv.org/abs/math-ph/0010025}.

\bibitem[{\citenamefont{Vermaseren}(2006)}]{Ver06}
\bibinfo{author}{\bibfnamefont{J.~A.~M.} \bibnamefont{Vermaseren}},
  \bibinfo{journal}{Nuclear Instruments and Methods in Physics Research A}
  \textbf{\bibinfo{volume}{559}}, \bibinfo{pages}{1} (\bibinfo{year}{2006}).

\bibitem[{\citenamefont{Bender and Orszag}(1999)}]{BO99}
\bibinfo{author}{\bibfnamefont{C.~M.} \bibnamefont{Bender}} \bibnamefont{and}
  \bibinfo{author}{\bibfnamefont{S.~A.} \bibnamefont{Orszag}},
  \emph{\bibinfo{title}{Advanced Mathematical Methods for Scientists and
  Engineers: Asymptotic Methods and Perturbation Theory}}
  (\bibinfo{publisher}{Springer}, \bibinfo{address}{New York},
  \bibinfo{year}{1999}).

\end{thebibliography}

\end{document}